\documentstyle[12pt]{article}
\begin{document}
\thispagestyle{empty}

\newcommand{\p}[1]{(\ref{#1})}
\newcommand{\be}{\begin{equation}}
\newcommand{\ee}{\end{equation}}
\newcommand{\sect}[1]{\setcounter{equation}{0}\section{#1}}
\renewcommand{\theequation}{\thesection.\arabic{equation}}
\newcommand{\vs}[1]{\rule[- #1 mm]{0mm}{#1 mm}}
\newcommand{\hs}[1]{\hspace{#1mm}}
\newcommand{\mb}[1]{\hs{5}\mbox{#1}\hs{5}}
\newcommand{\Db}{{\overline D}}
\newcommand{\bea}{\begin{eqnarray}}
\newcommand{\eea}{\end{eqnarray}}
\newcommand{\wt}[1]{\widetilde{#1}}
\newcommand{\und}[1]{\underline{#1}}
\newcommand{\ov}[1]{\overline{#1}}
\newcommand{\sm}[2]{\frac{\mbox{\footnotesize #1}\vs{-2}}
		   {\vs{-2}\mbox{\footnotesize #2}}}
\newcommand{\prt}{\partial}
\newcommand{\eps}{\epsilon}

\newcommand{\R}{\mbox{\rule{0.2mm}{2.8mm}\hspace{-1.5mm} R}}
\newcommand{\Z}{Z\hspace{-2mm}Z}

\newcommand{\cd}{{\cal D}}
\newcommand{\cg}{{\cal G}}
\newcommand{\ck}{{\cal K}}
\newcommand{\cw}{{\cal W}}

\newcommand{\vj}{\vec{J}}
\newcommand{\vl}{\vec{\lambda}}
\newcommand{\vz}{\vec{\sigma}}
\newcommand{\vt}{\vec{\tau}}
\newcommand{\vw}{\vec{W}}
\newcommand{\poiss}{\stackrel{\otimes}{,}}


\newcommand{\NP}[1]{Nucl.\ Phys.\ {\bf #1}}
\newcommand{\PL}[1]{Phys.\ Lett.\ {\bf #1}}
\newcommand{\NC}[1]{Nuovo Cimento {\bf #1}}
\newcommand{\CMP}[1]{Comm.\ Math.\ Phys.\ {\bf #1}}
\newcommand{\PR}[1]{Phys.\ Rev.\ {\bf #1}}
\newcommand{\PRL}[1]{Phys.\ Rev.\ Lett.\ {\bf #1}}
\newcommand{\MPL}[1]{Mod.\ Phys.\ Lett.\ {\bf #1}}
\newcommand{\BLMS}[1]{Bull.\ London Math.\ Soc.\ {\bf #1}}
\newcommand{\IJMP}[1]{Int.\ Jour.\ of\ Mod.\ Phys.\ {\bf #1}}
\newcommand{\JMP}[1]{Jour.\ of\ Math.\ Phys.\ {\bf #1}}
\newcommand{\LMP}[1]{Lett.\ in\ Math.\ Phys.\ {\bf #1}}

\renewcommand{\thefootnote}{\fnsymbol{footnote}}
\newpage
\setcounter{page}{0}
\pagestyle{empty}
\begin{flushright}
{April 1996}\\
{SISSA 56/96/EP}\\
{JINR E2-96-138}\\
{ hep-th/9604165}
\end{flushright}
\vs{8}
\begin{center}
{\LARGE {\bf Towards the construction of N=2 }}\\[0.6cm]
{\LARGE {\bf supersymmetric integrable hierarchies}}\\[1cm]

\vs{8}

{\large L. Bonora$^{a,1}$, S. Krivonos$^{b,2}$ and A. Sorin$^{b,3}$}
{}~\\
\quad \\
{\em ~$~^{(a)}$ International School for Advanced Studies (SISSA/ISAS),}\\
{\em Via Beirut 2, 34014 Trieste, Italy}\\
{\em {~$~^{(b)}$ Bogoliubov Laboratory of Theoretical Physics, JINR,}}\\
{\em 141980 Dubna, Moscow Region, Russia}~\quad\\

\end{center}
\vs{8}

\centerline{ {\bf Abstract}}
\vs{4}

We formulate a conjecture for the three different Lax operators that
describe the bosonic sectors of the three possible
$N=2$ supersymmetric integrable hierarchies
with $N=2$ super $W_n$ second hamiltonian structure. We check this
conjecture in the simplest cases, then we verify it in general
in one of the three possible supersymmetric extensions. To this
end we construct the $N=2$ supersymmetric extensions of the Generalized
Non-Linear Schr\"{o}dinger hierarchy by exhibiting the corresponding
super Lax operator. To find the correct hamiltonians we are led to a new
definition of super-residues for degenerate N=2 supersymmetric
pseudodifferential operators.
We have found a new non-polinomial Miura-like realization for $N=2$
superconformal algebra in terms of two bosonic chiral--anti--chiral free
superfields.
\vfill
{\em E-Mail:\\
1) bonora@frodo.sissa.it\\
2) krivonos@thsun1.jinr.dubna.su\\
3) sorin@thsun1.jinr.dubna.su }
\newpage
\pagestyle{plain}
\renewcommand{\thefootnote}{\arabic{footnote}}
\setcounter{footnote}{0}

\section{Introduction}

The construction of $N=2$ supersymmetric integrable  hierarchies is in 
progress. The motivations for studying them are diverse. On the one hand 
we have quite a good acquaintance with purely bosonic hierarchies and their
connection with physical models (2D and topological field theories), while
our knowledge of  $N=2$ supersymmetric integrable hierarchies is still 
scanty; in this regard the situation is quite different from conformal 
field theories, where  a great deal of attention has been paid to the 
$N=2$ supersymmetric extensions. On the other hand, once our understanding 
of $N=2$ supersymmetric integrable hierarchies becomes satisfactory, 
we may hope to find a link with physical models, e.g. with untwisted 
$N=2$ conformal field theories, in the wake of the relation between 
bosonic integrable hierarchies and topological field theories. A motivation 
of a different kind is the mathematical problem itself, which is 
challenging and  not devoid of surprises, such as, for example, 
the existence of three different $N=2$ extensions.
\par
In this letter we present a conjecture, and support it with many examples,
for the $N=2$ supersymmetric extensions of bosonic hierarchies
based on $W_n$ algebras. This conjecture is based on results obtained
in \cite{bon1,bon15,bon2} concerning the so--called $(n,m)$--KdV hierarchies.
Since both in
the latter and in the $N=2$ superextensions a $U(1)$ current plays a crucial
role, one may suspect that they might have much in common. This seems
to be the case.

The paper is organized as follows. In section 2 we introduce our conjecture
and explain how to make use of it. In section 3 we show that this conjecture
holds true in the $N=2$ superextensions of 2-- and 3--KdV. In section 4
we introduce the supersymmetric Lax operator for the generalized NLS
hierarchy and in section 5
we show that its bosonic limit reproduces one of the cases predicted
by our conjecture. In section 6 we discuss the problem of the transformation
between different pictures or bases, which are necessary in order to 
efficiently describe $N=2$ superextensions.

\section{The bosonic limit of $N=2$ supersymmetric integrable hierarchies
with $N=2$ super $W_n$ second hamiltonian structure}

Let us first summarize the state of affairs concerning N=2 superextensions.
Given a bosonic integrable hierarchy characterized by a certain number
of fields, say an $n$-KdV hierarchy for definiteness,
the corresponding N=2 superextension contains additional bosonic fields.
Therefore, while the final aim is to construct for any such superextension
the appropriate Lax operator, the first question one has to face
is what bosonic fields (obeying specific flow equations) we need to add to
have a system admitting $N=2$ supersymmetrization\footnote{Strictly speaking
one may find many
extensions: here we will consider only the minimal $N=2$ superextensions,
i.e. superextensions characterized by a minimal number of fields.}.
One of the possible ways to answer this question is to construct the
$N=2$ supersymmetric extension of the second hamiltonian structure of the
bosonic system, as a first step in the construction of the entire system.
Just in this way the $N=2$ supersymmetric KdV equation \cite{mat},
the $N=2$ Boussinesq  equation
\cite{ik2,y1}, $N=3,4$ KdV  equations \cite{bik,di} and generalized
KdV equation with $N=2$ super $W_4$ algebra as second Hamiltonian
structure \cite{y2}, were found. Up to now these are the cases in which
the N=2 superalgebra has been explicitly constructed, although an analogous
construction for general $W_n$ is only a matter of calculation.

In this approach a certain amount of
guesswork is necessary. Starting from the known second hamiltonian
structure one writes down a general hamiltonian with free parameters and
corresponding equations of motion. In all known cases
integrability of the constructed $N=2$ supersymmetric system holds
only for some special values of these parameters. It is interesting
that for the $N=2$ super KdV \cite{mat}, $N=2$ super--Boussinesq \cite{ik2,y1}
and $N=2$ extended 4--KdV equation \cite{y2} which possess $N=2$ $W_2,W_3$
and $W_4$ superalgebras as their second Hamiltonian structure, respectively,
there are three values of the parameters giving rise to three integrable
hierarchies in each case.

From the above it is clear that the bosonic sector of an N=2 superextension
contains a good deal of information concerning the supersymmetric system.
In this section we concentrate on the structure of bosonic sectors or limits
(all fermionic fields are set to zero) of the $N=2$ supersymmetric hierarchies
with $N=2$ super $W_n$ algebras as their second Hamiltonian structure.
For each $n$ we construct three different bosonic systems and three different
Lax operators, each of them leading to different integrable hierarchies. Then
we formulate our {\it Conjecture} about the structure of the Lax operators
for the bosonic sectors of
the $N=2$ supersymmetric integrable hierarchies with $N=2$ super-$W_n$ second
hamiltonian structure and, finally, specify the method for fixing the arbitrary
parameters in their $N=2$ hamiltonians.

The $N=2$ super $W_{m+1}$ algebras can be obtained via hamiltonian
reduction of the affine $sl(m+1|m)$ superalgebras with the set of
constraints corresponding to
the principal embedding of $sl(2|1)$ into $sl(m+1|m)$ \cite{pope}.
By construction, the bosonic limit of the $N=2$ $W_{m+1}$
superalgebra is represented in a suitable basis by the direct sum algebra
\be
W_{m+1}\oplus W_m\oplus U(1)  \label{boslim}
\ee
with the following relation among central charges
\be
\frac{c_{W_{m+1}}}{c_{W_m}}=-\frac{m+2}{m-1}, \; (m \geq 2). \label{crel}
\ee
Thus, after putting to zero all fermionic fields in the 
$N=2$ super-(m+1)-KdV
equations (with $N=2$ $W_{m+1}$ superalgebra as its second hamiltonian
structure), the bosonic equations will possess the algebra \p{boslim} as
second hamiltonian structure\footnote{The appearance of the graded structure
of the $sl(m|n)$ algebra (its Cartan part) in the second hamiltonian
structure of multi boson realization of KP hierarchy have been discussed
in the papers \cite{yu,ar}.}. The question we address here is how many
bosonic Lax operators providing the hamiltonian structure \p{boslim}
can be constructed. The key to answer this question is provided by the
analysis of the $(n,m)$-th KdV hierarchies of refs.\cite{bon1,bon15,bon2}.

Let us introduce the following n-KdV Lax operator
\be
L_{[n;\alpha ]}=\left( \alpha\partial \right)^n+
  \sum_{i=2}^{n} W_{i}^{(n)}(z) \left( \alpha\partial\right)^{n-i}
         \label{sorlax1}
\ee
where $\alpha$ is a numerical parameter. This Lax operator
induces, as second Hamiltonian structure on the currents
$W_{i}^{(n)}(z)$,  the $W_n$ algebra with central charge
\be
c_{W_n}=n(n^2-1)\alpha^2 \; . \label{cc1}
\ee
On the other hand, it was shown in \cite{bon2} that the Lax operator
$L_{[n,m;\alpha]}$
constructed from the $(n+m)$-KdV and the $m$-KdV Lax operators \p{sorlax1}
together with a $U(1)$ current $J(z)$, and defined by
$$
L_{[n,m;\alpha ]}  =  \left( e^{\frac{m}{2}\partial^{-1}(J(z))}
L_{[n+m;\alpha ]}
    e^{-\frac{m}{2}\partial^{-1}(J(z))}\right)\left(
   e^{\frac{m+n}{2}\partial^{-1}(J(z))}L_{[m;\alpha ]}
 e^{-\frac{m+n}{2}\partial^{-1}(J(z))}\right)^{-1}
$$
\be
  \equiv  \left( \nabla^{n+m}_{(m)}+\sum_{i=2}^{n+m}
    W_{i}^{(n+m)}(z)\nabla_{(m)}^{n+m-i} \right)
         \left( \nabla^{m}_{(n+m)}-\sum_{i=2}^{m}
    W_{i}^{(m)}(z)\nabla_{(n+m)}^{m-i} \right)^{-1} , \label{LL1}
\ee
where $\nabla_{(m)}$ is the covariant derivative
\be
\nabla_{(m)}\equiv \alpha \left( \partial -\frac{m}{2}J(z)\right)
\ee
gives rise to the second hamiltonian structure
\bea
W_{n+m}\oplus W_m \oplus U(1), && \mbox{ if } m\geq 2;  \label{hll}  \\
W_{n+1} \oplus U(1) ,&& \mbox{ if } m=1;  \label{hlla}  \\
W_{n}, && \mbox{ if } m=0 .\label{hllb}
\eea
These algebras have central charges
\be
c_{W_{n+m}}=(n+m)\left[ (n+m)^2-1\right]\alpha^2 ,\quad
c_{W_m}=-m(m^2-1)\alpha^2 \;.      \label{cc}
\ee
Let us notice, in particular, the identity
\be
 L_{[n,0;\alpha ]}\equiv L_{[n;\alpha ]}. \label{id}
\ee
The flow equations and hamiltonians are defined in the standard way
\bea
\frac{\partial}{\partial t_r} L_{[n,m;\alpha ]} & = &
A\left[ \left( L_{[n,m;\alpha ]}^{\frac{r}{n}} \right)_{+},
     L_{[n,m;\alpha ]}\right],  \\
\frac{\partial}{\partial t_r} W_i & = & \left\{ H_r^{(n,m)},W_i\right\},
                     \\
  H_r^{(n,m)} & \equiv & \int dx \mbox{ res}\left(
                 L^{\frac{r}{n}}_{[n,m;\alpha ]}\right) , \label{res0}
\eea
where $A$ is a normalization constant.
The subscript $(+)$ means differential part of pseudo--differential
operator and the usual definition of residue as the coefficient of
$\partial^{-1}$ is understood.

Now, using the Lax operators \p{LL1} and keeping in the mind
\p{hll}-\p{hllb} and the identity \p{id}
we can construct three Lax operators, which are our candidates to reproduce
the bosonic limit \p{boslim} of the $N=2$ $W_{m+1}$
superalgebra\footnote{The forth
possible Lax operator $L_{[m;\alpha ]}^{(4)}\equiv L_{[m+1;\alpha ]}\oplus
L_{[m-1,1;i\cdot \alpha]} \Rightarrow W_{m+1}\oplus \left( W_m \oplus U(1)
\right)$
with $m\geq 2$ produce systems which can not be recognize as bosonic
limit of $N=2$ hierarchies (at least for $m=2$).}
\begin{eqnarray}
L_{[m;\alpha ]}^{(1)}\equiv L_{[1,m;\alpha ]} & \Rightarrow &
             W_{m+1}\oplus W_m \oplus U(1) ,\label{con1} \\
L_{[m;\alpha ]}^{(2)}\equiv L_{[m,1;\alpha ]}\oplus  L_{[m;i\cdot\alpha ]}&
      \Rightarrow &
       \left( W_{m+1}\oplus U(1)\right) \oplus W_m , \label{con2} \\
L_{[m;\alpha ]}^{(3)}\equiv L_{[m+1;\alpha ]}\oplus
         L^M_{[m+1;i\cdot\alpha]} & \Rightarrow &
       W_{m+1}\oplus \left( W_m \oplus U(1) \right) , \label{con3}
\end{eqnarray}
where `$i$' is the imaginary unit and the
sign $\oplus $ between Lax operators in \p{con2}, \p{con3} means
that they act in different subspaces give rising to non-interacting systems.
The Lax operator
\be
L^M_{[n;\alpha ]}\equiv \nabla_{(2(n-1))}\left( \nabla^{n-1}_{(-2)}+
  \sum_{i=2}^{n-1}W_{i}^{(n-1)}(z)\nabla_{(-2)}^{n-i-1}\right)  \label{rem}
\ee
induces the $W_n$ algebra realized in terms of $W_{n-1}$ and $U(1)$
currents, see \cite{das}. One can check that the relation \p{crel}
between the central
charges of $N=2$ $W_{m+1}$ algebra is satisfied for the algebras
\p{con1}-\p{con3} with the corresponding central charges
\p{cc1}, \p{cc}.
Let us remark that for the special value $m=1$ the Lax operator
$L^{(1)}_{[1;\alpha ]}$ is equivalent to $L^{(2)}_{[1;\alpha ]}$.

Now we are ready to formulate our promised {\it Conjecture}:
{\it the bosonic sectors of the $N=2$
supersymmetric hierarchies with $N=2$ $W_{m+1}$ second hamiltonian structure
can be described by the Lax operators
\p{con1}-\p{con3}; correspondingly we will have three different $N=2$
hierarchies}.

We will substantiate this {\it Conjecture} later on with several
examples and by
constructing the super Lax operator corresponding to (\ref{con1}).
For the time being let us remark that an
interesting consequence of our {\it Conjecture} is the possibility to
construct the superfield hamiltonians for the $N=2$ supersymmetric
hierarchies almost straightforwardly. Indeed, if we write the most general
superfield expression for such hamiltonians in terms of the initial
supercurrents, which form the $N=2$ super $W_{m+1}$ algebra, and
put all fermionic fields equal to zero, then,
after passing to the components in the basis \p{boslim}, we can fix all the
coefficients by simply comparing the resulting expression with the
corresponding bosonic hamiltonian \p{res0}  following from the Lax operators
\p{con1}-\p{con3}.

Let us finally stress that among the Lax operators \p{con1}-\p{con3} only 
the first is `irreducible',  the remaining two being represented by the 
direct sum of two Lax operators. This means that, for the hierarchies induced 
by \p{con2} and \p{con3}, the fields belonging to different
irreducible addenda do not interact (they interact only via the fermionic
fields in the $N=2$ supersymmetric hierarchy). Thus in the bosonic limit, 
only the first
bosonic system with Lax operator \p{con1} is non trivial. This system is just
the $(1,n)$-KdV hierarchy\cite{bon2}. We will construct the $N=2$ 
supersymmetric
Lax operator for the $N=2$ supersymmetric extension of this system a bit
later after showing, in the next section, that our {\it Conjecture} 
works in the cases of the $N=2$ $W_3$ and $N=2$ $W_2$ hierarchies.

\setcounter{equation}0
\section{Examples}

In this section we will show how it is possible to reconstruct
the $N=2$ supersymmetric hamiltonians for $N=2$ $W_3$ and $W_2$ hierarchies
from their bosonic limits defined via our {\it Conjecture}.

\subsection{The N=2 supersymmetric Boussinesq hierarchies}

The $N=2$ super Boussinesq equation for which the second hamiltonian
structure is given by the classical $N=2$ super-$W_3$ algebra
\cite{pope}, has been constructed in \cite{ik1,y1}.
It can be defined as the system of two $N=2$
superfield equations for the supercurrents $J(Z), T(Z)$ having superspins
equal to $1,2$, respectively
\be
\frac{\partial T}{\partial t}=\left\{ T,H_2 \right\} ,\quad
  \frac{\partial J}{\partial t}=\left\{ J,H_2 \right\}
\ee
with the hamiltonian $H_2$
\be
H_2=\int dZ \left( T+a J^2 \right) ,\label{h2}
\ee
where $Z=(z,\theta,\bar\theta)$ is coordinate of $N=2$ superspace,
$dZ=dzd\theta d{\bar \theta}$ and $D,{\overline D}$ are the $N=2$
supersymmetric fermionic covariant derivatives
\begin{equation}\label{DD}
D=\frac{\partial}{\partial\theta}
 -\frac{1}{2}\bar\theta\frac{\partial}{\partial z} \quad , \quad
{\overline D}=\frac{\partial}{\partial\bar\theta}
 -\frac{1}{2}\theta\frac{\partial}{\partial z} ,
\end{equation}
$$
\left\{ D,{\overline D} \right\}= -\frac{\partial}{\partial z} \quad , \quad
\left\{ D,D \right\} = \left\{ {\overline D},{\overline D} \right\}= 0.
$$
The Poisson brackets between $J,T$ are defined to be $N=2$ super-$W_3$
algebra. Explicitly, the $N=2$ super Boussinesq equation reads
as follows
\begin{eqnarray}
\frac{\partial T}{\partial t} & = &
                  -2J''' -\left[ D,\Db \right]T' +\frac{80}{c}
   \left( \Db J DJ\right)' +\frac{32}{c}J'\left[ D,\Db \right] J
   +\frac{16}{c} J\left[ D,\Db \right] J' + \frac{256}{c^2}J^2J'
          \nonumber \\
 & + & \left( \frac{40}{c}-2 a\right)\Db J D T +
  \left( \frac{40}{c}-2 a \right) D J \Db T +
  \left( \frac{64}{c}+4 a \right) J'T+
  \left( \frac{24}{c}+2 a \right)JT' ,\nonumber \\
\frac{\partial J}{\partial t} & = &
    2T'- a \left( \frac{c}{4}\left[ D,\Db \right]J' -
        4JJ'\right) ,  \label{SB}
\end{eqnarray}
where $c=c_{W_3}$ is the central charge of the $N=2$ super $W_3$ algebra.
The integrability properties of the equation \p{SB} have been studied in
\cite{ik2,y1} where it was shown that the first six conserved currents exist 
for the following three values of the parameter $a$
\be
a^{(1)}=20/c,\quad a^{(2)}= -16/c,\quad a^{(3)}= -4/c .\label{pa}
\ee
However the Lax operator has been found only for the value $a^{(3)}$.

Let us demonstrate how our {\it Conjecture} can be applied to the $N=2$
super Boussinesq equation. From the second Hamiltonian structure of the 
bosonic sector and the Lax operators \p{con1}-\p{con3} we can
immediately write down the expressions for the bosonic
second order hamiltonian densities ${\cal H}_2$ \p{res0}
\begin{eqnarray}
{\cal H}_2^{(1)} & = & u_2-\frac{48}{c}u_1J_B-\frac{720}{c^2}J^3_B-
       \frac{72}{c} v_1J_B ,\label{1} \\
{\cal H}_2^{(2)} & = &
         u_2 +\frac{24}{c}u_1J_B-\frac{576}{c^2}J_B^3 , \label{2} \\
{\cal H}_2^{(3)} & = &
 u_2 -\frac{24}{c}v_1 J_B -\frac{144}{c^2}J_B^3 , \label{3}
\end{eqnarray}
where the currents $u_2,u_1$ form the $W_3$ algebra
\bea
\left\{ u_1(z_1),u_1(z_2)\right\} & = & \left( -\frac{c}{12}\partial^3+
  u_1\partial+\partial u_1\right) \delta (z_1-z_2) ,\nonumber \\
\left\{ u_1(z_1),u_2(z_2)\right\} & = & \left( \frac{c}{24}\partial^4+
  u_2\partial+2\partial u_2-\partial^2 u_1\right) \delta (z_1-z_2) ,\nonumber \\
\left\{ u_2(z_1),u_2(z_2)\right\} & = & \left( \frac{c}{36}\partial^5+
  \partial^2 u_2-u_2\partial^2  +\frac{16}{c}u_1\partial u_1-
  \frac{2}{3}u_1\partial^3-\frac{2}{3}\partial^3 u_1\right) \delta (z_1-z_2) ,
       \label{39}
\eea
while $v_1$ and $J_B$ span $W_2\oplus U(1)$
\bea
\left\{ v_1(z_1),v_1(z_2)\right\} & = & \left( \frac{c}{48}\partial^3+
  v_1\partial+\partial v_1\right) \delta (z_1-z_2) ,\nonumber \\
\left\{ J_B(z_1),J_B(z_2)\right\} & = &  -\frac{c}{36} \delta' (z_1-z_2) ,
 \label{310}
\eea
where derivatives and currents appearing in the right hand side
are understood to be evaluated at $z_1$.
We have fixed the normalization of the hamiltonians by setting equal to 1
the coefficient of the spin 3 current $u_2$. Let us note that it is
straightforward to pass from the currents in terms of which the Lax
operators \p{con1}-\p{con3} are defined, to the currents $u_2,u_1,v_1,J_B$.
For example for the  Lax operator $L^{(2)}_{[2;\alpha]}$ \p{con2}, which
can be represented in the following equivalent form
\be
L^{(2)}_{[2;\alpha]}=
 \left(\alpha^2 \partial^2+w_1+w_2\frac{1}{\alpha\partial-S_1}\right)\oplus
  \left(-\alpha^2 \partial^2 +w_0 \right),
\ee
these transformations (up to an automorphism $J_B\Rightarrow -J_B$,
$u_2\Rightarrow -u_2+u_1'$, $u_1\Rightarrow u_1$ of the algebras
\p{39}, \p{310}) are
\bea
w_0 & = & v_1 ,\qquad\quad S_1=-\frac{3}{2}J_B , \nonumber \\
w_1 & = & u_1-\frac{18}{c}J_B^2-\frac{3}{2}J_B' , \nonumber \\
w_2 & = & u_2+\frac{24}{c}u_1J_B-\frac{576}{c^2}J_B^3-
    \frac{72}{c}J_BJ_B'-J_B'' .
\eea
To compare the bosonic limit of the $N=2$ superfield hamiltonian \p{h2}
with \p{1}-\p{3}, we must first integrate over $\theta,\bar\theta$ and
put all fermionic fields equal to zero
\be
H_2\Rightarrow H_2^{B}=\int dz \left( \left[ D,\Db \right] T|+
      2 a J| \left[ D,\Db \right] J| \right),
\ee
where $|$ denotes the $\theta,\bar\theta$ independent part (i.e. the limit
$\theta =  \bar\theta = 0 $), and then pass
to the bosonic components $u_1,u_2,v_1,J_B$ which form the
$W_3\oplus W_2 \oplus U(1)$ \p{39}, \p{310}.
In the case at hand the correspondence is
(up to the above mentioned automorphism of the algebras \p{39}, \p{310})

\begin{eqnarray}
J| & = & 3J_B , \nonumber \\
\left[ D, \Db \right] J| & = & 2u_1+2v_1-\frac{36}{c}J_B^2 , \nonumber \\
T| & = & -u_1-4v_1 , \nonumber \\
\left[ D,\Db \right] T| & = & -6u_2+3u_1'+\frac{48}{c}(u_1+4v_1)J_B .
           \label{tr}
\end{eqnarray}
Using this we get the following expression for $H_2^B$
\be
H_2^B=-6\int dz \left( u_2-\frac{8}{c}\left( 1+\frac{a c}{4}\right)
  u_1J_B -\frac{16}{c}\left( 2+ \frac{a c}{8}\right) v_1J_B+
  \frac{36 a}{c} J_B^3  \right) . \label{hbb}
\ee
It is easy to check that the ${\cal H}_2^B$ \p{hbb} coincides (up to
an inessential rescaling) with ${\cal H}_2^{(1)},
{\cal H}_2^{(2)}$, ${\cal H}_2^{(3)}$ \p{1}-\p{3}, exactly for the values of
parameter $a$ \p{pa}, respectively.

We stress that our {\it Conjecture} and the procedure we have proposed
do not guaranteed the integrability and the conservation of the hamiltonians
we have constructed. It provides a necessary but not sufficient condition
for integrability. The safest and most
timesaving way to prove integrability is through explicit construction of
Lax operators.
Unfortunately this is very complicated even for the $N=2$ $W_3$ hierarchies.
The main problem is that the variables (supercurrents) in which the second
hamiltonian structure is transparent are not very suitable for writing
the Lax operators. As an example we present here the  possible Lax
operators for the $N=2$ super Boussinesq which correspond to the
following values of parameters: $a^{(1)} =20/c,a^{(2)}=-16/c$.

The starting point is the following general form of the Lax operators
\be
L=\partial + \kappa_0 + \sum_{i=1}^{\infty} \left( \kappa_i + \sigma_i\Db+
  {\bar\sigma}_iD +\rho_i\left[ D,\Db \right]\right)\partial^{-i} ,
   \label{lgen}
\ee
where $\kappa,\sigma,\bar\sigma $ and $\rho$ are sums of monomials
constructed from
supercurrents $J(Z), T(Z)$ with the proper dimensions. We will find out that
the flow equations can appear in the two different forms
\be
\frac{\partial}{\partial t_n} L = A\left[ \left( L^n \right)_{+},L\right]
           \label{le1}
\ee
or
\be
\frac{\partial}{\partial t_n} L = B\left[ \left( L^n\right)_{\geq 1},L\right],
           \label{le2}
\ee
where the subscript $(+)$ means differential part of a pseudodifferential
operator, while $(\geq 1)$ means the pure differential part with
exclusion of the
constant term. We do not know from the start which flow equations
we have to use in order to reproduce the equations \p{SB}, so we have to check 
all the possibilities. The results of our calculations can be summarized as
follows
\begin{center}
\underline{$a^{(1)} = 20/c$.}
\end{center}
\begin{eqnarray}
A=-1,\; L_{(1)} & = & \frac{c}{8}\partial+27\left[ \Db D  J\right]
    \partial^{-1}-27\left[ \Db D \left(
 U+\frac{36}{c}J^2-\frac{5}{2}J'\right)\right]\partial^{-2} \nonumber \\
 & + & \left[ \Db D \left( -54 U'+108 J''+
 \frac{1944}{c}JU+\frac{46656}{c^2}J^3 \right.\right.
           \nonumber \\
 & - & \left.\left. \frac{972}{c}JJ'+\frac{5832}{c}J\Db DJ\right)-
    \frac{5832}{c}\left[ \Db D J\right] \Db D J \right]\partial^{-3} +
               \ldots  ,\label{lx1}\\
A=-3, \; L_{(2)} & = & \frac{c}{8}
              \partial-3\left[ D\Db J\right]\partial^{-1}
      -  \left[ D\Db \left( U+\frac{36}{c}J^2-\frac{3}{2}J' \right)\right]
  \partial^{-2} +\ldots , \label{lx2}
\end{eqnarray}
where $U=T+\frac{1}{2}\left[ D,\Db \right] J-\frac{16}{c}J^2$ and
the flow equations are the \p{le1}. (The square brackets mean that the
derivatives act only on the terms inside brackets.)

\begin{center}
\underline{$a^{(2)}= -16/c$.}
\end{center}
\bea
B=-3,\;L_{(1)}  & = & \frac{c}{8} \partial +  \frac{3}{2}J+
   \frac{3}{2}\Db \partial^{-1}
  \left[ DJ\right] -\frac{1}{2}\partial^{-1}{\tilde U}
   \nonumber \\
 & + &
  \left( \frac{1}{2}\left[ D{\tilde U}\right] \Db + \frac{6}{c}J{\tilde U}+
  \frac{1}{4}\left[ \left[ D,\Db \right] {\tilde U}\right] -
  \frac{9}{c}\left[ DJ\right] \left[ \Db J\right] \right)\partial^{-2} +
          \ldots , \label{lx3}
\eea
\be
B=-3,\; L_{(2)}  = \frac{c}{8} \partial -\frac{3}{2}DJ\Db\partial^{-1}+
 \frac{1}{2}D\left( {\tilde U}+\frac{3}{2}\left[ \left[ D,\Db\right] J\right]+
  \frac{3}{2}J'\right) \Db\partial^{-2}+\ldots , \label{lx4}
\ee
where ${\tilde U}= T+\frac{1}{2}\left[ D,\Db\right] J+\frac{2}{c}J^2$
and the Lax equation is given by \p{le2}.

Above we have reported only the first few terms of the Lax operators,
but we have checked
them as far as the $\partial^{-5}$ terms. We did not succeed in finding 
a closed form for these Lax operators. On the other hand in the next
section we will present the Lax operator for the case $a^{(1)}=20/c$, using
new variables which renders its structure more transparent.

In the rest of this section we discuss the problem of extracting the
hamiltonians from the Lax operators \p{lx1}-\p{lx4}.

First of all, let us note that the Lax operator \p{lx4} contains
the standard residue, i.e. the
coefficient before $[D,{\overline D}]\partial^{-1}$ \cite{mat}.
Therefore in this case the hamiltonians can be written down as
\be
H_n=\int dZ \mbox{ res}\left( L^n \right) . \label{resid4}
\ee

The Lax operator \p{lx3} does not have the
standard residue (which vanishes). We can in such a case apply the
definition
of residues as the constant part of the pseudo-differential operator, see
\cite{krivsor2}. We checked that the first four hamiltonians can
be constructed following this definition, i.e.
\be
H_n=\int dZ \left( L^n \right)_0 , \label{resid3}
\ee

As for the Lax operators \p{lx1}, \p{lx2},  they
do not contain neither the latter nor the standard $N=2$ residues \cite{mat}.
Moreover, these Lax operators look like pure
bosonic ones, since they do not contain spinor $D$ or $\Db$
operators acting separately.
It is very interesting that the corresponding hamiltonians
(at least the first four ones) can be obtained in these cases using the
definition of residues for bosonic pseudodifferential operators, i.e. as
the integrated coefficients of $\partial^{-1}$
\be
H_n=\int dz \mbox{ res}_{B}\left( L^n \right), \label{resid12}
\ee
where the integration is over the space coordinate $z$. The main reason
why this definition works is that the coefficients before
$\partial^{-1}$ in $L^n$  \p{lx1}-\p{lx2},
at least as far as $n=4$, can be represented as
\be
 \mbox{ res}_{B}\left( L^n \right) =
\left[ D,\Db \right]  \left( {\cal H}_n \right) +
\mbox{ full space derivative terms } .
\ee

The proof that our definitions of hamiltonians, or equivalently
of the so--called residues, work in the case of general
degenerated $N=2$ pseudodifferential operators, for which the standard
residues are equal to zero, is an open problem which goes beyond the scope
of this paper. However the above explicit calculations seem to indicate
that proper alternative residues can always be defined.

\subsection{The N=2 supersymmetric KdV hierarchies}

The $N=2$ super KdV equation
\be
{\dot J}= -J''' +\frac{36}{c}\left( J\left[ D,\Db\right] J\right)'-
 \frac{ac+12}{2c}\left( \left[ D,\Db \right] J^2\right)'+
  \frac{36a}{c}J^2J'   \label{kdv}
\ee
has been constructed in \cite{mat} starting from the Hamiltonian
\be
H_3=\int dZ \left( J\left[ D,\Db \right]J + \frac{a}{3} J^3 \right)
      \label{hkdv}
\ee
and the Poisson brackets between the supercurrent $J(Z)$ are supposed to be
$N=2$ superconformal algebra \cite{adem}
\be
\left\{ J(Z_1),J(Z_2) \right\} = \left( -\frac{c}{24} \left[ D,\Db
\right]\partial +\Db J D+ DJ\Db +J\partial+ \partial J \right)
 \delta (Z_1-Z_2) , \label{n2sca}
\ee
where
$$\delta (Z_1-Z_2)=(\theta_1-\theta_2 )({\bar \theta}_1-
  {\bar \theta}_2 ) \delta (z_1-z_2)$$
is the delta function in $N=2$ superspace, and derivatives and supercurrents
appearing in the right hand side are understood to be evaluated at $Z_2$.
This equation is proved to be integrable for
the following values of parameter $a$ \cite{mat,popow3}
\be
a^{(1)}=a^{(2)}=-48/c,\; a^{(3)}=24/c,\; a^{(0)}=-12/c .
\ee
Then the three bosonic Lax operators \p{con1}-\p{con3}
reduce to two independent ones (see discussion below formula \p{rem}).
Following the same arguments as for $N=2$ super $W_3$,
we can find the third hamiltonians for these bosonic hierarchies in terms
of the currents $v_1, J_B$ forming the $W_2\oplus U(1)$ algebra \p{310}
\bea
{\cal H}^{(1)}_3 & = & {\cal H}^{(2)}_3=v_1^2-\frac{108}{c}v_1J_B^2-
   \frac{3}{4} J_BJ_B''+\frac{1620}{c^2}J_B^4 ,\label{ha1} \\
{\cal H}^{(3)}_3 & = & v_1^2-\frac{3}{4}J_BJ_B''-
                    \frac{324}{c^2}J_B^4 . \label{ha2}
\end{eqnarray}

Now, as in the case of the $N=2$ super Boussinesq equation, we integrate
over $\theta,\bar\theta $ in \p{hkdv} and pass to the
components $v_1,J_B$
\be
v_1=-\frac{1}{2}\left[ D,\Db \right] J|-\frac{6}{c}J^2| ,
                             \quad J_B=\frac{i}{\sqrt{3}}J|  \label{nv}
\ee
with the Poisson brackets \p{310}. It is easy to check that the bosonic
limit of \p{hkdv}, written in terms of $v_1,J_B$ \p{nv},
\be
{\cal H}_3^B=4 \left(v_1^2+\frac{3}{2}\left(a -\frac{24}{c} \right) v_1J_B^2
-\frac{3}{4}J_BJ_B''-
  \frac{27}{c}\left( a-\frac{12}{c}\right) J_B^4\right)
\ee
coincides with \p{ha1}, \p{ha2} (up to an inessential rescaling)
for $a^{(1)}=a^{(2)}=-48/c $ and $a^{(3)}=24/c$, respectively.
The case $a^{(0)} = -12/c$ is exceptional and cannot be described
whithin the scheme of our {\it Conjecture} (this case has been discussed
in \cite{popow3}).

To close this section let us stress that
the described procedure is certainly applicable not only to the first
meaningful hamiltonians but also to the few subsequent ones.
We can retrieve the
hamiltonians from the bosonic parts for which the Lax operators
(and thus the hamiltonians) are defined by
\p{con1}-\p{con3}. Based on our {\it Conjecture} we can also claim that for
the $N=2$ hierarchies
with bosonic limit given by \p{con3} there are no supersymmetric hamiltonians
of order $k(m+1), k=1,2,\ldots $ because they do not exist in the bosonic case
($H_n=\int dz \mbox{ res }((L^{(3)}_{[m+1,\alpha ]})^{n/(m+1)})=0$ for
$n=k(m+1)$ since $L^{(3)}_{[m+1,\alpha ]}$ is a pure differential operator).
We have checked this for all known six hamiltonians in $N=2$ $W_2,W_3$ and $W_4$
cases. However our procedure is expected to fail if
the hamiltonian under investigation contains superfield terms
which after integration over $\theta,\bar\theta$  disappear in the bosonic
limit. The coefficients of such terms cannot be fixed in the framework of our
{\it Conjecture}. The first appearance of such
terms might take place in $H_{8}$ ( $\int dZ DJ\Db J DJ'\Db J'$ term) but
it is unclear now whether such terms do occur in hamiltonians.
It might be that the coefficients of such terms vanish in hamiltonians 
of $N=2$ supersymmetric integrable hierarchies.
If this is true, one can reconstruct all hamiltonians from their bosonic limit.

\setcounter{equation}0
\section{The N=2 supersymmetric GNLS hierarchies}

In this section we construct the $N=2$ supersymmetric generalization of the
bosonic generalized non--linear Schr${\rm \ddot o}$dinger
(GNLS) hierarchies and their manifestly $N=2$ supersymmetric
Lax operators. In the following section we will recognize that their bosonic
limits coincide with the hierarchies defined by the Lax operators \p{con1}
(in fact a direct sum of two of them).

Let us introduce $(n+m)$  pairs of chiral and anti--chiral $N=2$
superfields $F_A(Z)$ and ${\overline F}_A(Z)$ with capital Latin indices
$A,B=1,\ldots ,n+m$
\be
D F_A(Z)= {\overline D}\; {\overline F}_A(Z) = 0 \; ,
\label{chiral}
\ee
which are fermionic for $A=1,\ldots ,n$ and bosonic for
$A=n+1,\ldots ,n+m$. The grading is $F_AF_B=(-1)^{d_Ad_B}F_BF_A$, where
$d_A=1 (d_A=0)$ for fermionic (bosonic) superfields.
In what follows, we find it convenient to denote the indices of the
fermionic superfields by small Latin letters
$i,j=1,\ldots , n$ and the bosonic ones by Greek letters
$\alpha,\beta=1,\ldots , m$, respectively\footnote{For simplicity
we call bosonic (fermionic) the commuting (anticommuting) fields. However
they may not have the usual spin--statistics connection. In the latter case 
we can perhaps identify them with BRST ghosts.}.

Our ansatz for the Lax operator of the $N=2$ super GNLS hierarchy is
\begin{eqnarray}
L&=& \partial - \frac{1}{2} F_A(Z) {\overline F}_A(Z)  -
 \frac{1}{2} F_A(Z) {\overline D}
\partial^{-1} \left[ D {\overline F}_A(Z)\right] \; ,
\label{suplax}
\end{eqnarray}
where summation over repeated indices is understood and the square
brackets mean that the fermionic derivative $D$ act only on the term
${\overline F}_A$ inside brackets. Such operator
provides (for integer $k$) the consistent flows
\begin{eqnarray}
{\textstyle{\partial\over\partial t_k}}L &=&[ (L^k)_{\geq 1} , L]  \; ,
\label{laxfl1}
\end{eqnarray}
where the subscript $\geq 1$ means that only the purely derivative
part must be considered. Let us note that only the spin of the product 
$F_A {\overline F}_A$ is fixed by the ansatz \p{suplax},
and it is equal to $1$ .

The first flow from \p{laxfl1} is trivial
\begin{eqnarray}
{\textstyle{\partial\over \partial t_1}} F_A &=& F_A' \; , \nonumber\\
{\textstyle{\partial\over\partial t_1}} {\overline F}_A &=&
{\overline F}_A' \; ,
\end{eqnarray}
while the second reads
\begin{eqnarray}
{\textstyle{\partial\over\partial t_2}} F_A&=&
F_A'' +  {D}(F_B {\overline F}_B\;\! {\overline D} F_A),
\nonumber\\
{\textstyle{\partial\over \partial t_2}} {\overline F}_A &=&
-{\overline F}_A'' + {\overline D}
(F_B {\overline F}_B {D}{\overline F}_A) \; . \label{supgnls}
\end{eqnarray}
Beside global $N=2$ supersymmetry the Lax operator \p{suplax} and
the flows \p{laxfl1} are invariant with respect to the $GL(n|m)$
supergroup. Let us note that the second flow \p{supgnls} in the case
$n=1,m=0$, which corresponds to one pair of fermionic
chiral--anti--chiral superfields, is just the $N=2$ super NLS
equation \cite{krisor1}.
The Lax operator \p{suplax} generalizes to 
the multi-components case (i.e., for arbitrary values of $n$ and $m$)
the Lax operator for $N=2$ super NLS hierarchy \cite{krivsor2}.
Moreover, as we will show in the next section, the bosonic limit of these
systems coincides with bosonic GNLS hierarchies.
That is why we call the corresponding
hierarchies the $N=2$ super GNLS hierarchies.

Let us write down the explicit expressions for the
first three hamiltonians
\begin{eqnarray}
 H_1 &=& -\frac{1}{2}\int dZ 
\left( F_A {\overline F}_A \right) \; ,\nonumber\\
 H_2 &=& \int dZ \left( F_A {\overline F}_A' +
\frac{1}{4} (F_A {\overline F}_A)^2 \right) \; , \label{supham} \\
 H_3 &=& - \frac{3}{2} \int dZ \left( F_A {\overline F}_A'' -
\frac{1}{2} {\overline D} (F_A {\overline F}_A)
\cdot D (F_B {\overline F}_B) +
F_A {\overline F}_A' \cdot F_B {\overline F}_B +
\frac{1}{12} (F_A {\overline F}_A)^3 \right) . \nonumber
\end{eqnarray}
It is interesting to remark that the first hamiltonian density
${\cal H}_1$
satisfy the following equation of motion
\be
\frac{\partial}{\partial t_2} {\cal H}_1 = ({\cal H}_1'+{\cal H}_2)'\; ,
\ee
so one can find the additional integral of motion
\be
{\tilde H}_1 = -\frac{1}{2}\int dz \left( F_A {\overline F}_A \right) \; ,
\ee
where we have only space integration.

Note that the standard definition of residue
in the case of $N=2$ supersymmetric pseudodifferential operators as the
coefficient of $[D,{\overline D}]\partial^{-1}$, once again cannot be
applied to our Lax operator \p{suplax}, due to vanishing of such
residues for any power of $L$. Nevertheless, the infinite number
of conserved currents can be extracted from $L$ \p{suplax} as follows
\begin{eqnarray}
{H}_k = \int d Z (L^k)_{0} \; , \label{res}
\end{eqnarray}
where the subscript $0$ means the constant part of the operator.
We do not have a general proof that for an arbitrary degenerate
$N=2$ supersymmetric pseudodifferential operator the constant part
gives the correct hamiltonians, but for the case at hand
this statement has been explicitly checked for the hamiltonians \p{supham}.

It is instructive to rewrite the second flow equations \p{supgnls} in terms
of unconstrained $N=1$ superfields
${\cal F}_A({\tilde Z}), {\overline {\cal F}}_A({\tilde Z})$,
where ${\tilde Z}=(z,{\theta}_2)$
is the coordinate of the $N=1$ real superspace. This can be done by solving
the chirality conditions \p{chiral} via superfields
${\cal F}_A({\tilde Z}), {\overline {\cal F}}_A({\tilde Z})$
\begin{eqnarray}
F_A(Z) &=& {\cal F}_A({\tilde Z}) + i {\theta}_1 D_2 {\cal F}_A({\tilde Z}) ,
\nonumber\\
{\overline F}_A(Z) &=& {\overline {\cal F}}_A({\tilde Z})
- i {\theta}_1 D_2 {\overline {\cal F}}_A({\tilde Z}) \; . \label{n1var}
\end{eqnarray}
Here ${\theta}_1\equiv \frac{\theta + \bar\theta}{2},
{\theta}_2\equiv \frac{\theta - \bar\theta }{2i}$ are real Grassmann
coordinates and $N=1$ supersymmetric fermionic covariant derivatives
$D_1, D_2$ are defined by
\begin{equation}\label{DD1}
D_1\equiv D+{\overline D}=\frac{\partial}{\partial{\theta}_1}
 -{\theta}_1\frac{\partial}{\partial z} \quad , \quad
D_2\equiv i \left(D-{\overline D}\right)=\frac{\partial}{\partial{\theta}_2}
 -{\theta}_2\frac{\partial}{\partial z} ,
\label{covder}
\end{equation}
$$
D_1^{2}=D_2^{2}= -\frac{\partial}{\partial z} \quad , \quad
\left\{ D_1,D_2 \right\} =  0.
$$
Substituting \p{n1var} into \p{supgnls} and using \p{covder} we get
the following second flow equations
\begin{eqnarray}
{\textstyle{\partial\over\partial t_2}} {\cal F}_A&=&
{\cal F}_A'' +
(-1)^{d_B} {\cal F}_B D_2 ({\overline {\cal F}}_B\;\! D_2 {\cal F}_A),
\nonumber\\
{\textstyle{\partial\over \partial t_2}} {\overline {\cal F}}_A &=&
-{\overline {\cal F}}_A'' +
{\overline {\cal F}}_B D_2 ({\cal F}_B\;\! D_2 {\overline {\cal F}}_A)
\; . \label{n1supgnls}
\end{eqnarray}
The systems \p{n1supgnls} still possess $N=2$
supersymmetry. However the latter is hidden due to the description by means of
$N=1$ superfields. The simplest case with $n=1, m=0$ has been 
considered in \cite{roe,das1} and
called there the $N=1$ NLS equation. Now, generalizing the Lax operator
for $N=1$ NLS \cite{das1}, we can construct
the Lax operator in terms of $N=1$ superfields for the system \p{n1supgnls}
\be
L = \partial - \frac{1}{2} {\cal F}_A({\tilde Z})
{\overline {\cal F}}_A({\tilde Z})  -
 \frac{1}{2} {\cal F}_A({\tilde Z}) D_2 \partial^{-1}
\left[ D_2 {\overline {\cal F}}_A({\tilde Z})\right] \; .
\label{n1suplax}
\ee
This representation in terms of $N=1$ superfields might be more convenient
to analyze hamiltonian reduction from affine $N=1$ superalgebras.

\setcounter{equation}0
\section{Bosonic limit of the N=2 supersymmetric GNLS hierarchies}

In the $N=2$ supersymmetric extended KP hierarchies we may select two main
different bases (i.e., choices of elementary fields). It was first 
recognized in
\cite{krisor1,krivsor2} that while in one basis the second Hamiltonian
structure of $N=2$ supersymmetric extensions is transparent, in the second
one the Lax operator can be written in an elegant and economic way.
In the previous section we constructed the $N=2$ supersymmetric
extension of the GNLS hierarchies in the basis where their
Lax operator structure is clear. However it is not evident to which
bosonic systems we considered in section 2 these equations correspond.
In this section we consider the bosonic limit of $N=2$ hierarchies
with the Lax operators \p{suplax} and establish their relations
with the $(1,n)$-KdV hierarchies.

To consider the bosonic limit of the second flow equations
\p{supgnls}, let us define the components of the fermionic superfields as
\be
f_i = \frac{1}{\sqrt{2}} {\overline D} F_i| \; , \;
{\overline f}_i = \frac{1}{\sqrt{2}} D {\overline F}_i| \; , \;
{\psi}_i  =  \frac{1}{\sqrt{2}} F_i| \; , \;
{\overline {\psi}}_i =  \frac{1}{\sqrt{2}} {\overline F}_i|
\ee
and the components of the bosonic superfields as
\be
 {\xi}_{\alpha} =  \frac{1}{\sqrt{2}} {\overline D} F_{\alpha}| \; , \;
 {\overline {\xi}}_{\alpha} =
        \frac{1}{\sqrt{2}} D {\overline F}_{\alpha}| \; , \;
b_{\alpha}  = \frac{1}{\sqrt{2}} F_{\alpha}| \; , \;
{\overline b}_{\alpha} = \frac{1}{\sqrt{2}} {\overline F}_{\alpha}| \quad ,
\ee
where $|$ means the $({\theta}, {\bar\theta})\rightarrow 0$
limit. So, $\psi_i,{\overline \psi}_i,\xi_{\alpha},
{\overline \xi}_{\alpha}$ are
fermionic fields while $f_i,{\overline f}_i,b_{\alpha},
{\overline b}_{\alpha}$ are bosonic ones.

In terms of such components the equations \p{supgnls} become
\bea
\frac{\partial}{\partial t_2} \left( \begin{array}{c} \psi_i \\ b_{\alpha}
\end{array} \right) & = &  \left( \begin{array}{c} \psi_i \\ b_{\alpha}
\end{array} \right)''+2\left( -\psi_j{\bar
f}_j+b_{\beta}{\bar\xi}_{\beta}\right)
\left( \begin{array}{c} f_i \\ b_{\alpha}
\end{array} \right)-2\left(\psi_j\bar \psi_j+b_{\beta}\bar b_{\beta}\right)
\left( \begin{array}{c} \psi_i \\ b_{\alpha}
\end{array} \right)',\nonumber \\
\frac{\partial}{\partial t_2} \left( \begin{array}{c} f_i \\ \xi_{\alpha}
\end{array} \right)& = & \left( \begin{array}{c} f_i \\ \xi_{\alpha}
\end{array} \right)''+2\left( -f_j{\bar f}_j-\psi_j{\bar \psi}_j'-
 b_{\beta}{\bar b}_{\beta}'+\xi_{\beta}{\bar\xi}_{\beta}\right)
\left( \begin{array}{c} f_i \\ \xi_{\alpha}
\end{array} \right) \label{comp} \\
 & -&
 2\left( \psi_j{\bar\psi}_j+b_{\beta}{\bar b}_{\beta}\right)
\left( \begin{array}{c} f_i \\ \xi_{\alpha}
\end{array} \right)'  +  2\left( -\psi_j{\bar f}_j-{\bar
\psi}_jf_j+b_{\beta}{\bar\xi}_{\beta}-{\bar b}_{\beta}\xi_{\beta}\right)
\left( \begin{array}{c} \psi_i \\ b_{\alpha}
\end{array} \right)' .\nonumber
\eea
The equations for the components ${\bar\psi}_i,{\bar f}_i,
{\bar b}_{\alpha},{\bar\xi}_{\alpha}$ can be obtained from \p{comp}
by conjugation ($t_2$ is pure imaginary and under complex conjugation
$t_2^{*}=-t_2$).

To get the bosonic limit we have to put all the fermionic
fields ${\psi}_i, {\overline {\psi}}_i, {\xi}_{\alpha},
{\overline {\xi}}_{\alpha}$ to zero. This leaves us with the following set
of bosonic equations
\begin{eqnarray}
{\textstyle{\partial\over\partial t_2}} f_i&=&
f_i'' -  2 f_i f_j {\overline f}_j - 2 b_{\beta}
  ({\overline b}_{\beta} f_i)',
\nonumber\\
{\textstyle{\partial\over \partial t_2}} {\overline f}_i &=&
-{\overline f}_i'' + 2 {\overline f}_i f_j {\overline f}_j -
2 {\overline b}_{\beta} (b_{\beta} {\overline f_i})'   \; ,
\label{bosgnls1}
\end{eqnarray}
\begin{eqnarray}
{\textstyle{\partial\over\partial t_2}} b_{\alpha}&=&
b_{\alpha}'' -  2 b_{\beta} {\overline b}_{\beta} b_{\alpha}',
\nonumber\\
{\textstyle{\partial\over \partial t_2}} {\overline b}_{\alpha} &=&
-{\overline b}_{\alpha}'' - 2 b_{\beta}
   {\overline b}_{\beta} {\overline b}_{\alpha}'   \; .
\label{newgnls1}
\end{eqnarray}

Let us note that in \p{bosgnls1}, \p{newgnls1} only the spins of the products
$f_j {\overline f}_j$ and $b_{\beta} {\overline b}_{\beta}$
are fixed and they are equal to 2 and 1, respectively.

After passing to the new fields $g_i, {\overline g}_i$ defined by
\begin{eqnarray}
g_i & = & f_i {\exp (-{\partial^{-1}} (b_{\beta} {\overline b}_{\beta}))},
     \nonumber\\
{\overline g}_i & = &{\overline f}_i { \exp ({\partial^{-1}}
(b_{\beta} {\overline b}_{\beta})) } \label{newgnls}
\end{eqnarray}
we can rewrite the equations \p{bosgnls1} as
\begin{eqnarray}
{\textstyle{\partial\over\partial t_2}} g_i&=&
g_i'' -  2 g_i g_j {\overline g}_j,
\nonumber\\
{\textstyle{\partial\over \partial t_2}} {\overline g}_i &=&
-{\overline g}_i'' + 2 {\overline g}_i g_j {\overline g}_j \; .
\label{bosgnls2}
\end{eqnarray}
The fields  $b_{\alpha}, {\overline b}_{\alpha}$
are completely decoupled and
the set of equations \p{bosgnls2} form the
GNLS equations \cite{fk}. They can be viewed as
the second flow of  GNLS hierarchies with the
Lax operators ${\tilde L}_1$
\begin{eqnarray}
{\tilde L}_1&=& \partial - g_i \partial^{-1} {\overline g}_i
\label{boslax1}
\end{eqnarray}
and the flow equations
\begin{eqnarray}
{\textstyle{\partial\over\partial t_k}}L &=&[ (L^k)_{+} , L] \; .
\label{laxfl2}
\end{eqnarray}

As for the $b_{\alpha},{\overline b}_{\alpha}$
fields, which obey the closed set of equations \p{newgnls1},
after passing to the new fields $r_{\alpha},{\overline r}_{\alpha}$
\begin{eqnarray}
r_{\alpha}&=& b_{\alpha}'
{\exp (-{\partial^{-1}} (b_{\beta} {\overline b}_{\beta}))},
\nonumber\\
{\overline r}_{\alpha}&=& {\overline b}_{\alpha}
{\exp ({\partial^{-1}} (b_{\beta} {\overline b}_{\beta}))} \; ,
\label{newgnls2}
\end{eqnarray}
we get the following system
\begin{eqnarray}
{\textstyle{\partial\over\partial t_2}} r_{\alpha}&=&
r_{\alpha}'' -  2 r_{\alpha} r_{\beta} {\overline r}_{\beta},
\nonumber\\
{\textstyle{\partial\over \partial t_2}} {\overline r}_{\alpha} &=&
-{\overline r}_{\alpha}'' +
2 {\overline r}_{\alpha} r_{\beta} {\overline r}_{\beta} \; ,
\label{bosgnls3}
\end{eqnarray}
which coincides with the GNLS equations. We remark that the transformations
\p{newgnls2} look like formulas for `bosonization of bosons'.

Thus the set of equations \p{bosgnls1}, \p{newgnls1}, which represent the
bosonic limit of the $N=2$ supersymmetric equations \p{supgnls},
can be converted to the form \p{bosgnls2}, \p{bosgnls3} which is
nothing but the direct sum of two GNLS systems \cite{fk}.
But for  this system, the transformations to the
$(1,n)$-KdV hierarchies  we considered in section 2
are well known \cite{ar2,bon2}. Thus we can conclude that the $N=2$
supersymmetric hierarchies with Lax operators  \p{suplax} are the $N=2$
supersymmetric extensions of the direct sum of
$(1,n)$-KdV and $(1,m)$-KdV hierarchies with the Lax operator
$L^{(1)}_{[n;\alpha ]}\oplus L^{(1)}_{[m;\alpha ]}$ \p{con1}.
For the special cases with only bosonic ($n=0$) or fermionic ($m=0$)
superfields in the Lax operator \p{suplax} we have supersymmetric
extensions of $(1,m)$- or $(1,n)$-KdV hierarchies, respectively.
Thus there are two possibilities to construct $N=2$ supersymmetric
extension of $(1,n)$-KdV hierarchies, based on bosonic or fermionic
superfields. One expects that these superextensions be
somehow related. As an example, in the next section we will estimate
the explicit relation between
$N=2$ supersymmetric extensions of the $(1,1)$-KdV hierarchy with fermionic
and bosonic superfields.

Let us point out that, despite the existence of
the transformations \p{newgnls2}, that bring the systems
of equations for
$b_{\alpha},{\overline b}_{\alpha}$ \p{newgnls1}
to the  GLNS system \p{bosgnls3},
the \p{newgnls1} system
seems to be more fundamental. In fact the system \p{newgnls1} possesses
the conserved current
\be
{\cal H}_0 = b_{\alpha} {\overline b}_{\alpha} \label{bosham}
\ee
which is missing for the \p{bosgnls3} system. For completeness we
present here the Lax operator for the $b_{\alpha},{\overline b}_{\alpha}$
system
\begin{eqnarray}
{\tilde L}_2&=& \partial -
b_{\alpha}' (\partial +
b_{\beta} {\overline b}_{\beta} )^{-1} {\overline b}_{\alpha}.
\label{boslax2}
\end{eqnarray}

Finally we would like to stress that an alternative
approach to the construction of $N=2$ supersymmetric hierarchies
recently proposed in \cite{popow1} seems to be less general than ours.
The author uses general superfields (instead of chiral ones, \p{chiral})
and a completely different ansatz for the super Lax operators
(instead of ours, \p{suplax}).
The absence of a chirality condition leads to a $4(n+m)$ component
system in the bosonic limit (instead of a $2(n+m)$ one, as in our case).
As a consequence the systems constructed in this and the previous sections for
odd values of $n+m=2l-1, l=1,2,\ldots $, cannot be reproduced within the
approach of \cite{popow1}. For the other cases, one should prove that
the bosonic limits do reproduce GLNS hierarchies, also beyond
the case $n=1, m=0$, which is the only one proved in
\cite{popow1}. If this is true, it would be interesting
to find the relation with our $N=2$ supersymmetric hierarchies
(for even values of $n+m$).

\setcounter{equation}0
\section{Transformations between bases}

In the previous sections we showed that there are two preferred
bases in which $N=2$ supersymmetric extensions of $(1,n)$-KdV
hierarchies
can be constructed. In one basis the second hamiltonian structure is
transparent and it coincides with $N=2$ super $W_n$ algebra, while
the structure of Lax operators is complicated. In the second basis the
Lax operators can be written in the explicit form \p{suplax}, but the second
hamiltonian structure is hidden. For the bosonic limit,
when all fermionic fields equal zero, the transformations
between these bases are known \cite{ar2}. But in the case of $N=2$
supersymmetric extensions these transformations are more complicated.
In this section we present some preliminary results concerning these
transformations for the $N=2$ super-KdV and super-Boussinesq
hierarchies.

\subsection{N=2 super-KdV case ($a^{(1)}=-48/c$)}

The $N=2$ super-KdV equation ($N=2$ $(1,1)$-KdV)
has the $N=2$ superconformal algebra \p{n2sca}
as the second hamiltonian structure \cite{mat}, and corresponds to one pair
of bosonic $B(Z),{\bar B}(Z)$ $(n=0,m=1)$
or fermionic $F(Z),{\bar F}(Z)$ $(n=1,m=0)$ chiral-anti-chiral superfields
in the second basis with Lax operator given by \p{suplax}.
The transformation from $F(Z),{\bar F}(Z)$ to the single spin 1
$N=2$ superfield $J(Z)$, which obeys the standard $N=2$ super-KdV hierarchy
equations, has been found in \cite{krisor1}
\be
J=\frac{c}{24}\left( \frac{1}{2} F{\bar F} +
\frac{\partial}{\partial z} \left( \mbox{ln} {D {\bar F}}\right)
   \right) .  \label{kdvtr1}
\ee
It is interesting that the second hamiltonian structure for the
$F,\bar F$ fields coincides with $N=2$ super ${\hat U}(2)$ algebra and
the hamiltonians belong to the coset \cite{krivsor2}
$$
\frac{N=2 \mbox{ super } {\hat U}(2)}{U(1)\oplus U(1)} .
$$

In the second case ,when we deal with bosonic superfields 
$B(Z),{\bar B}(Z)$, the transformation to the superfield $J(Z)$ 
has the following form
\be
J=-\frac{c}{24}\left( \frac{1}{2}B{\bar B} +
     \frac{\partial}{\partial z}\mbox{ln} {\bar B}\right) .
   \label{kdvtr2}
\ee
It can be easily checked that the second hamiltonian structure for
the superfields $B(Z),{\bar B}(Z)$ is rather simple
\be
\left\{ B(Z_1){\bar B}(Z_2)\right\} = \frac{48}{c}D\Db \delta(Z_1-Z_2)
 \label{chirpb}
\ee
and is related to the $N=2$ superconformal algebra \p{n2sca} through the
transformation \p{kdvtr2}. Thus in the case of the bosonic superfields
$B(Z),{\bar B}(Z)$ the transformation \p{kdvtr2} is defined to be a Poisson
map relating
the Poisson structure \p{chirpb} and the $N=2$ superconformal algebra.
We would like to stress that the expression \p{kdvtr2} gives new
classical non-polynomial Miura-like realization
of $N=2$ superconformal algebra in terms of free bosonic chiral--anti--chiral
superfields $B(Z),{\bar B}(Z)$ with Poisson brackets \p{chirpb}.
It is very interesting to find the applications of the realization
\p{kdvtr2} to e.g. $N=2$ superstrings, superconformal field theories, etc.

Let us note that the transformations \p{kdvtr1}, \p{kdvtr2} are not
self-conjugate, but the conjugate ones are also allowed transformations.

To estimate the relation between two different descriptions of $N=2$
$(1,1)$-KdV hierarchy in terms of fermionic $F(Z),{\bar F}(Z)$ \p{kdvtr1}
and bosonic $B(Z),{\bar B}(Z)$ \p{kdvtr2} superfields one can equate
\p{kdvtr1} and \p{kdvtr2}. The resulting relation is
\be
{\bar B}D{\bar F}\mbox{ exp}\left( \frac{1}{2}\partial^{-1}
    \left( F{\bar F}+B{\bar B} \right) \right) = \mbox{const} \; .
\ee

\subsection{N=2 super-Boussinesq case ($a^{(1)}=20/c$)}

This second non trivial example of the $N=2$ $(1,n)$-KdV hierarchies
at $n=2$ corresponds
to the case of four chiral-anti-chiral superfields with Lax operator
\p{suplax}. There are
three different possibilities: a) all four superfields are
fermionic $(n=2,m=0)$ b)two superfields are fermionic and two bosonic
$(n=1,m=1)$ and
c) all superfields are bosonic $(n=0,m=2)$.
As we showed in the previous section, for the cases a) and c) the bosonic
limits  coincide with four component GNLS hierarchy. The equivalence
between GNLS and $(1,n)$-KdV hierarchies was proved in \cite{ar2}.
The simplest $N=2$ supersymmetric
extension of the bosonic transformations from the four component GLNS
to the $(1,2)$-KdV hierarchies look as follows
\bea
J&=&\frac{c}{24}\left[ \frac{1}{2}F_i{\bar F}_i+
\frac{\partial}{\partial z}\mbox{ln}\left(
 D{\bar F}_2D{\bar F}_1'-D{\bar F}_2'D{\bar F}_1+
 \kappa D\left( F_i{\bar F}_i\right) D\left( {\bar F}_1{\bar F}_2\right)
           \right) \right], \nonumber \\
T & = & -\frac{c}{16}\left[ \frac{1}{2}\left(F_i'{\bar F}_i-F_i{\bar
F}_i'-\frac{1}{2}(F_i{\bar F}_i)^2\right)-\frac{16}{c}\left( \frac{5}{2}
\left[ D,\Db \right] J-\frac{20}{c}J^2\right) \right.\nonumber \\
 & + & \left. \frac{\partial}{\partial t_2}\mbox{ln}\left(
 D{\bar F}_2D{\bar F}_1'-D{\bar F}_2'D{\bar F}_1+
 \kappa D\left( F_i{\bar F}_i\right) D\left( {\bar F}_1{\bar F}_2\right)
           \right) \right] ,
 \label{sbtr}
\eea
where $t_2$ is related to $t$ in equation \p{SB} by $t_2=-3t$ and
time derivatives of $F_i(Z),{\bar F}_j(Z)$ are defined by \p{supgnls}.
It can be checked that the  \p{sbtr} reproduces the
correct expressions for the bosonic components of the superfields $J(Z)$ and
$T(Z)$ \p{SB} in terms of
bosonic components of the fermionic superfields (case a)
$F_1,{\bar F}_1,F_2,{\bar F}_2$, see the transformations in \cite{ar2}.
But unfortunately \p{sbtr} fails to give the true transformations
if the fermionic components are included. This is a bit surprising, since
other terms which preserve the $GL(2)$ symmetry of the equations \p{supgnls}
for the
superfields $F_1,{\bar F}_1,F_2,{\bar F}_2$, and can be added to
\p{sbtr}, are non-local (the terms in the \p{sbtr} which are proportional
to the parameter $\kappa$ are the only local ones which possess 
$GL(2)$ symmetry and they disappear in the bosonic limit). As for case b),
the bosonic limit is given by the direct sum of two NLS equations
and so it cannot be related with the $(1,2)$-KdV bosonic hierarchy.
Thus the problem of superfield transformations
between the systems \p{suplax} we constructed in this paper
and the $N=2$ super $(1,n)$-KdV hierarchies with $N=2$ super $W_n$ algebra
second Hamiltonian structure, is still open.

\section{Conclusion}

In this paper we have presented the results of our study of the
possible $N=2$ supersymmetric integrable hierarchies with $N=2$
$W_n$ superalgebra as the second hamiltonian structure. We have presented 
our {\it Conjecture} about the Lax operators for their bosonic limits.
Using it one can reconstruct the supersymmetric hamiltonians.
We have also constructed
the $N=2$ supersymmetric extensions of bosonic GNLS hierarchies as
well as of $(1,n)\oplus (1,m)$-KdV hierarchies and their Lax operators in
terms of $N=2$ superfields.
We have found a new non-polynomial realization for the $N=2$ superconformal
algebra and proposed a new definition of super-residues for
degenerate $N=2$ supersymmetric pseudodifferential operators.
The transformation between different superfield representations has
been analized for the simplest cases. A detailed analysis of this
complicated problem is under way.

\section*{Acknowledgements}
Two of us (S.K. and A.S.) would like to thank M. Tonin and F. Toppan for
useful discussions.

S.K. and A.S. thank SISSA and the University of Padova for hospitality
during the course of this work. This investigation has been
supported in part by grants INTAS-93-633, INTAS-94-2317.

\end{document}